# Fabrication and Characterization of Superconducting Quantum Interference Device using $(Bi_{1-x}Sb_x)_2Se_3$ Topological Insulator Nanoribbons

Nam-Hee Kim[1], Hong-Seok Kim[1], Yiming Yang[2], Xingyue Peng[2], Dong Yu[2], Yong-Joo Doh[1*]

[1]Department of Physics and Photon Science, School of Physics and Chemistry, Gwangju Institute of Science and Technology (GIST), Gwangju 61005, Korea

[2]Department of Physics, University of California, Davis, CA 95616, U.S.A.

*E-mail: yjdoh@gist.ac.kr

**Abstract**

We report on the fabrication and electrical transport properties of superconducting quantum interference devices (SQUIDs) made from a $(Bi_{1-x}Sb_x)_2Se_3$ topological insulator (TI) nanoribbon (NR) connected with $Pb_{0.5}In_{0.5}$ superconducting electrodes. Below the transition temperature of the superconducting $Pb_{0.5}In_{0.5}$ electrodes, periodic oscillations of the critical current are observed in the TI NR SQUID under a magnetic field applied perpendicular to the plane owing to flux quantization. Also the output voltage modulates as a function of the external magnetic field. Moreover, the SQUID the SQUID shows a voltage modulation envelope, which is considered to represent the Fraunhofer-like patterns of each single junction. These properties of the TI NR SQUID would provide a useful method to explore Majorana fermions.

Topological insulators (TIs) are bulk insulators having metallic surface states that are topologically protected by time-reversal symmetry [1]. The topological surface states (TSSs) are not subject to backscattering by nonmagnetic impurities because the surface electrons are aligned perpendicular to their translational momentum by strong spin–orbit coupling [2]. This property leads to a highly quantum-coherent transport, making TIs candidates for quantum information devices [3].

When TIs are combined with conventional s-wave superconductors, Majorana fermions (MFs) can be expected at the interface [1]. At the transparent interface between the superconductor and the TI, supercurrent in topological phase can flow through the TI owing to the superconducting proximity effect [4]. An experimentally verified Josephson supercurrent was found within various TIs such as $Bi_2Se_3$ [5-7], $Bi_2Te_3$ [8, 9], and $Bi_{1.5}Sb_{0.5}Te_{1.7}Se_{1.3}$ [10]. When two superconducting junctions are connected in parallel in a device called a direct current superconducting quantum interference device (dc-SQUID), the device is very sensitive to a magnetic flux. Thus, it plays a crucial role in a superconducting flux qubit [11].

Quantum interference devices have received attention because of the charge neutrality and zero energy of the MF. MFs form zero energy Andreev bound states in the barrier when the phase difference between the superconducting leads is π. Therefore these quasiparticles are described by a current–phase relation (CPR) with 4π periodicity instead of 2π one in TI/superconductor system[12]. The CPR of SQUIDs shows more clear signatures than single Josephson junction because the quantum interference is determined by the relative phase difference between the junctions. The CPR has been studied in flake-shaped TI-based SQUIDs [8, 13].

In this paper, we present an experimental study of SQUIDs made of a TI that has a nanoribbon (NR) shape instead of a flake shape. Because NR shapes have a larger surface-to-volume ratio than flake shapes, a stronger surface effect can be expected. The surface states of TI NR is can be host a single gapless linearly dispersing Dirac fermion. The TI NR which has a quasi-1D geometry exhibits Majorana end modes when it is passed through the magnetic flux of half-integer flux quanta and proximitized from the superconductor [14]. The TI NR SQUID exhibited clear modulation of the critical current and output voltage as a function of the external magnetic field owing to superconducting phase interference effects. Furthermore, Fraunhofer-like patterns superimposed on the SQUID oscillations were observed. Our experimental study gives useful information for flux qubit applications based on TI NRs.

As-grown Sb-doped $Bi_2Se_3$ TI NRs were mechanically transferred to a highly n-doped silicon substrate covered by a 290-nm-thick oxide layer with prepatterned Ti/Au bonding pads. Source and drain electrodes were patterned by standard electron beam lithography followed by e-beam evaporation of 300-nm-thick $Pb_{0.5}In_{0.5}$. A 10-nm-thick Au film was used as a capping layer. The PbIn alloy source was made using a mixture of lead(Pb) and indium(In) pellets with a weight ratio of 1:1. Before metal deposition, the NR surface was cleaned by oxygen plasma treatment to remove e-beam resist residues. Then, the NR was treated using a 6:1 buffered oxide etch for 7 s to remove a native oxide layer that was assumed to be present. Figure 1 (a) shows a scanning electron microscopy (SEM) image of sample **D1** after the device fabrication process was complete. The PbIn superconducting electrodes exhibit a superconducting transition below critical temperature $T_c$ = 6.8 K and magnetic field $H_{c,perpendicular}$ = 0.74 T [15]. We measured the electrical transport properties of the TI NR SQUID using four-point measurement in a closed-cycle helium cryostat and $^3$He refrigerator (Cryogenic) down to base temperatures of 2.4 and 0.3 K, respectively. For low-noise measurement, two-stage resistor-capacitor (RC) filters (cutoff frequency = 10 kHz) and π filters were used in conjunction with the measurement leads at room temperature [16], and an additional one-stage RC filter was connected in series with the sample in a closed-cycle helium cryostat at 2.4 K. The junction properties are summarized in table 1.

**Table 1.** Physical parameters of the SQUIDs: $L_1$. $L_2$ are the channel length, and $w$ is the width of the SQUID junction. $L_{SQ}$ and $W_{SQ}$ are the length and width of the SQUID, respectively. Max. $I_c$ is a maximum critical current, $R_N$ is a normal resistance of each sample, and $T_{base}$ is a base temperature.

| Sample | $L_1$ (nm) | $L_2$ (nm) | $w$ (nm) | $L_{SQ}$ (μm) | $W_{SQ}$ (μm) | Max. $I_c$ (μA) | $I_c R_N$ (μV) | $T_{base}$ (K) |
|--------|-----------|-----------|---------|---------------|---------------|-----------------|----------------|----------------|
| D1 | 290 | 257 | 414 | 1.41 | 1.59 | 2.60 | 13.8 | 0.3 |
| D2 | 114 | 89 | 495 | 1.26 | 1.36 | 0.40 | 2.88 | 2.4 |
| D3 | 157 | 121 | 506 | 3.43 | 1.34 | 0.45 | 5.07 | 2.4 |
| D4 | 243 | 292 | 368 | 1.66 | 1.46 | 7.20 | 39.6 | 0.3 |
| D5 | 260 | 260 | 391 | 1.70 | 1.30 | 0.10 | 1.62 | 2.5 |

TIs have topologically protected surface states and thus are expected to show weak antilocalization (WAL) when a perpendicular magnetic field is applied to the sample. Figure 1 (b) shows angle-dependent differential magnetoconductivity (MC) [Δσ = σ(B) − σ(0)] curves obtained at $T = 3.1$ K from the same batch of SQUIDs. The observed MC curves show a sharp cusp-like shape near zero magnetic field, which is a typical feature of WAL, and merge into a single curve representing the normal component of the magnetic field. According to the Hikami–Larkin–Nagaoka (HLN) formula,

$$\Delta\sigma_{2D} = \alpha \frac{e^2}{2\pi^2\hbar^2}\left(ln\left(\frac{\hbar}{4eBL_\phi^2}\right) - \psi\left(\frac{1}{2} + \frac{\hbar}{4eBL_\phi^2}\right)\right), \tag{1}$$

where $e$ is the electron charge, $\hbar$ is the reduced Planck's constant, $L_\phi$ is the phase coherence length, $\psi(x)$ is the digamma function, $B$ is the out-of-plane applied magnetic field, and $\alpha$ is a dimensionless transport parameter [17]. Because the WAL effect is considered to be a distinctive transport property of the TSS, the relationship between the parameter $\alpha$ and the number of conducting channels plays an important role in characterizing the transport behavior in TIs [18]. Each conducting channel contributes 0.5 to $\alpha$, so $\alpha$ is expected to be 1 in the ideal two-dimensional (2D) system. In our experiments, however, $\alpha$ is estimated to be 1.31, and this can be considered as the origin of the 2D electron gas (2DEG) state on the top surface [19]. At the interface between a TI and another material, an energy change (band bending) is introduced into the electronic band owing to intrinsic surface/interface states, impurities, or applied electric fields. This band bending near the surface of the TI induces the formation of a 2DEG [20] that contributes a value of ~1 to the $\alpha$ value of the top surface ($\alpha_{top}$). However, the magnitude of the $\alpha$ value of the bottom surface, $\alpha_{bottom}$, decreases to ~0.2–0.3 because the WAL is reduced by scattering between the TSS and the 2DEG state [21]. As a result, the total $\alpha$ ($\alpha_{total} = \alpha_{top} + \alpha_{bottom}$) is ~1.2–1.3, which is consistent with our experimental data. In addition, the phase coherence length is estimated to be 465 nm from the HLN fitting result, which confirms the two-dimensionality of the system and is similar to previous results of TI studies [22, 23]. Consequently, we verify the 2D nature and presence of the TSS in the $(Bi_{1-x}Sb_x)_2Se_3$ TI NR in our device.

When an external magnetic field is applied along the length of the TI NR, periodic magnetoconductance oscillation is observed which is called Aharonov–Bohm oscillation (ABO) and has a period of $h/e$, where $h$ is Planck's constant. ABO is generally caused by the interference of partial waves formed in the closed electron path surrounding the magnetic flux in the mesoscopic ring structure [24]. However, for TIs, ABO is known to be caused by the one-dimensional (1D) subband in the TI surface state [25-27]. Figure 1 (c) shows the differential magnetoconductance curves under an axial magnetic field at $T = 2.4$ K of devices from the same batch of SQUIDs. A gate-induced shift between 0-ABO at $V_g = -16.7$ and $-17.1$ V and π-ABO at $V_g = -16.8$ V was observed. 0-ABO is an oscillation with a conductance peak at zero magnetic flux, and π-ABO shows a conductance peak at the half-flux quanta. The observed period of the magnetoconductance oscillation is $\Delta B = 1700$ Oe, which suggests that the effective area of the NR ($A_{eff} = 0.024$ μm$^2$) calculated using $h/e = \Delta B \cdot A_{eff}$ is comparable to the geometric area ($A = 0.032$ μm$^2$). The difference between the geometric area and the effective area is caused by an oxide layer on the TI NR surface. Because of the density-of-states effect of the 1D subband caused by surface electron conduction and the existence of a π Berry phase due to the TSS, the ABO that alternates between the 0 and π phases depending on the gate voltage can be measured far from the Dirac point [25-29]. We observed the alternating ABO near $V_g \sim -17$ V, which is distant from the Dirac point ($V_{Dirac} \sim -55$ V). We also observed oscillation with a period of $h/2e$, which is induced by the effect of WAL correction caused by breaking of time-reversal symmetry [27, 29].

The proximity effect of supercurrent though the (Bi$_{1-x}$Sb$_x$)$_2$Se$_3$ TI NR SQUID is observable below the $T_c$ value of the PbIn superconducting electrode. Figure 2 (a) shows the current–voltage (*I–V*) curves under a magnetic field applied perpendicular to the TI NR SQUID loop with increasing magnetic flux ($\Phi$) at $T = 2.4$ K. The critical current $I_c$, which indicates the proximity strength of two weak links, is the switching current at which a finite voltage is measured when switching from superconductive to dissipative conduction. The maximum $I_{c,max}$ (at zero flux) of device **D2** is 0.40 μA at $T = 2.4$ K. The critical current has a maximum value at $\Phi = 0$ and vanishes at $\Phi_0/2$, in agreement with the prediction that $I_c$ is modulated with the period of the magnetic flux quantum, $\Phi_0 = h/2e$, owing to interference resulting from the superconducting wavefunction split along the arms of both sides of the loop by the flux. The magnetic flux quantum has the relation $\Phi_0 = \Delta B \times S_{eff}$ for the effective area of the SQUID loop, $S_{eff}$ (the yellow dashed line in figure 1a), and the magnetic field periodicity ($\Delta B$) in a conventional SQUID. The effective area of the SQUID is calculated from the measured magnetic field period $\Delta B = 4.5$ Oe as 4.6 μm$^2$. On the other hand, the geometric area (inner area of the SQUID; see figure 1d) is 1.71 μm$^2$ because of the London penetration depth ($\lambda_L$) to which the magnetic field penetrates the superconducting electrode. Thus, $\lambda_L$ values of 0.57 and 0.56 μm are obtained for samples **D2** and **D4**, respectively, from

$S_{eff} = (2\lambda_L + W_{SQ}) \times (L_{SQ} + \lambda_L)$ (see figure 1 (c)), which are close to the values obtained from another nanowire-based SQUID [30].

The critical current shows a periodic modulation as a function of the magnetic flux in the flux quantization SQUID loop [4]. The periodic modulation of $I_c$ for sample **D2** is shown in a color plot of the differential resistance, $dV/dI$, in figure 2 (b). Here, the dark region is the superconducting state with $dV/dI = 0$, and the critical current is modulated with a period of 4.5 Oe. The periodic modulation of $I_c(\Phi)$ can be described by the sinusoidal CPR in the SQUID loop. The self-inductance of the SQUID loop is thought to be negligible; thus, $I_c(\Phi)$ is given by

$$I_c(\Phi) = [(I_{c1} - I_{c2})^2 + 4I_{c1}I_{c2}\cos^2(\pi\Phi/\Phi_0)]^{1/2} , \qquad (2)$$

where $I_{c1}$ and $I_{c2}$ are the critical currents of the weak links [31]. The critical currents of each weak link in sample **D2** are obtained as $I_{c1} = 204$ nA and $I_{c2} = 198$ nA by fitting (2) to the experimental data (see the white line in figure 2b). It is deduced that the two weak links in the TI NR SQUID are almost identical, with $I_{c2}/I_{c1} = 0.97$. The self-inductance is estimated to be $L_s \sim 3.5$ pH from the SQUID geometry, which is consistent with the screening parameter $\beta_L = 2\pi L_s I_0/\Phi_0 = 2.2 \times 10^{-3} \ll 1$, where $I_0 = (I_{c1} + I_{c2})/2$.

A decrease in the critical current as the magnetic flux increased from 0 to $1/2\Phi_0$ was similarly measured in sample **D4** at a much lower temperature, $T = 0.3$ K, as shown in figure 2 (c). At zero flux, $I_{c,max}$ is obtained 7.2 µA but there is a finite critical current ($I_{c,min}$) at $\Phi = \Phi_0/2$ instead of the supercurrent "off" state ($I_c = 0$). The absence of the $I_c$-off state can be considered to indicate asymmetry of the critical current through the two weak links [32], where $I_{c1}$ and $I_{c2}$ are evaluated as $I_{c1} = (I_{c,max} + I_{c,min})/2$ and $I_{c2} = (I_{c,max} - I_{c,min})/2$. Consequently, $I_{c1} = 4.8$ µA, $I_{c2} = 2.4$ µA, and $I_{c2}/I_{c1} = 0.50$ for sample **D4**. Figure 2 (d) shows a color plot of the $dI/dV$ vs. $\Phi$ for sample **D4**. The critical current oscillated periodically between $I_{c,max}$ and $I_{c,min}$ with the flux, as in sample **D2**, and a finite supercurrent at $\Phi = (n + 1/2)\Phi_0$, where $n$ is an integer, was observed, as expected.

When a constant current bias ($I_{DC}$) is applied to the SQUID, the output voltage $V$ is related to the applied flux. The voltage modulation for the TI NR SQUID (sample **D2**) as a flux-to-voltage transducer is shown in figure 3 (a). The output voltage of the SQUID is sinusoidally modulated with the magnetic flux. As $I_{DC}$ increases, the oscillation amplitude of the voltage increases and becomes maximum near $I_c$, which is similar to the result of a previous study using a $Bi_2Te_3$ flake SQUID [8]. The maximum value of the voltage oscillation is related to the sensitivity of the SQUID, $|\partial V/\partial \Phi|$, and the sensitivity of sample **D2** is shown in figure 3 (b). A maximum $|\partial V/\partial \Phi|$ value of 17 µV/Φ is obtained near $I_c$.

The output voltage of sample **D4**, which has an asymmetric critical current, is also related to the oscillation amplitude of $I_{DC}$ and shows a maximum value near $I_c$, as shown in figure 3 (c). In this case, the voltage oscillation is not completely sinusoidal but has a similar tendency. Figure 3 (d) displays the SQUID sensitivity of sample **D4**; a maximum sensitivity of 152 µV/$\Phi_0$ is obtained, which is 10 times higher than that of sample **D2**. These results can be considered to reveal a difference in the $I_c R_N$ values of each sample, which are 2.9 µV for **D2** and 40 µV for **D4**. These are higher than those reported in previous works on a TI SQUID, 1.6 µV at $T = 0.3$ K [7] and 34.1 µV at $T = 0.01$ K [13]. The critical current at zero flux of sample **D4** is 7.2 µA, whereas that of sample **D2** is 16 times smaller ($I_c = 0.4$ µA). Because the proximity effect of the superconductor is inversely proportional to the thermal energy, the lower temperature of sample **D4** leads to the higher critical current, and the single junction length is longer (see table 1).

Figure 4 (a) shows an additional interference pattern envelope in the higher magnetic field regime that was superposed on the SQUID oscillation. The envelope originates from Fraunhofer-like interference due to each Josephson junction [32]. The Fraunhofer pattern appears when the dimension of a single Josephson junction is in the wide junction limit, $w \gg \xi_B$, where the magnetic length $\xi_B = \sqrt{\Phi_0/B}$. The magnetic length of sample **D2**, $\xi_B = 224$ nm, is shorter than the width of a single junction ($w = 495$ nm). The observed periods are 96 and 79 Oe at the first and second envelopes, respectively; these values are similar to the period calculated from the dimension of a single weak junction (82 Oe). A period of 4.5 Oe is obtained for the SQUID oscillation observed in figure 4a, which is in agreement with the value measured from the critical current.

The magnetic field modulation of the critical current is plotted in figure 4 (b). When the Fraunhofer diffraction effect is added to the SQUID, the CPR can be described by the following formula [31, 32]:

$$I_c(B) = [(I_{c1} - I_{c2})^2 + 4I_{c1}I_{c2}\cos^2(\pi\Phi/\Phi_0)]^{1/2} \left|\sin\left(\frac{\pi\Phi_J}{\Phi_0}\right)/\left(\frac{\pi\Phi_J}{\Phi_0}\right)\right|, \qquad (3)$$

where $I_{c1}$ and $I_{c2}$ are the critical currents for each junction, and $\Phi$ is the magnetic flux through the square-ring area of the SQUID. $I_{c1}$ and $I_{c2}$ values of 204 and 198 nA, respectively, are obtained, which are similar to the results for the conventional SQUID CPR, and the envelope of the first interference pattern has a period of 92 Oe. This result is consistent with the value that was verified in figure 4a. Simultaneous measurement of the Fraunhofer diffraction pattern in the SQUID was also observed in a previous study using a $Bi_2Te_3$ flake SQUID [9]. In this case, the period of the SQUID oscillation corresponded to a smaller area than the geometric area. In our experiments, the TI NR SQUID oscillation period is comparable to the effective area considering the penetration depth.

In conclusion, we demonstrated the realization of a $(Bi_{1-x}Sb_x)_2Se_3$ TI NR SQUID in contact with PbIn superconducting electrodes. We observed critical current modulation under a magnetic field applied perpendicular to the plane of the SQUID. The output voltage also oscillated with the magnetic field, so the device acted as a flux-to-voltage transducer with high SQUID sensitivity. In addition, a Fraunhofer-like pattern was superimposed on the SQUID interference pattern. Our results are expected to be useful for investigating the superconducting phase interference effects and MFs.

**Figure Captions**

Figure 1 | **a.** SEM image of SQUID sample **D1**. Yellow dashed line shows the effective area of the supercurrent loop. **b.** Angle-dependent differential MC [$\Delta\sigma = \sigma(B) - \sigma(0)$] of another device that used an NR from the same batch at $T = 3.1$ K. Solid line is best fit to (1). **c.** Differential magnetoconductance curve for different gate voltages $V_g$ of −16.7, −16.8, and −17.1 V (bottom to top) of devices that used an NR from the same batch at $T = 2.4$ K. Dashed lines are guides for the eye. **d.** Schematic of SQUID and single Josephson junction. $L_{SQ}$ is the length of the SQUID, $W_{SQ}$ is the width of the SQUID, and $\lambda_L$ is the penetration depth. The effective Josephson junction length $L_{JJ}^{eff}$ is the sum of the junction lengths of each side ($L_1$, $L_2$) and 2 times the penetration depth.

Figure 2 | **a.** Current–voltage (*I–V*) curves for different magnetic fluxes through the TI NR SQUID loop at $T = 2.4$ K, where $\Phi/\Phi_0 = 0$ (black), 0.1 (red), 0.2 (blue), 0.3 (green), 0.4 (magenta), and 0.5 (purple) (sample **D2**). **b.** Color plot of d$V$/d$I$ as a function of dc current, magnetic field, and flux. Black region indicates the supercurrent regime; white line is the CPR using (2). **c.** *I–V* curves at different magnetic fluxes of sample **D4** ($T = 0.3$ K). **d.** Color plot of d$V$/d$I$ as a function of dc current, magnetic field, and flux.

Figure 3 | **a.** Modulation of output voltage as a function of $\Phi$ for sample **D2** at $T = 2.4$ K. The bias current $I_{DC}$ increases from 0.2 to 1.0 µA from bottom to top. **b.** Corresponding maximum values of $|\partial V/\partial \Phi|$ versus the biasing current *I* of sample **D2**. Error bars are indicated. **c.** Modulation of output voltage as a function of $\Phi$ for sample **D4** at $T = 0.3$ K. The bias current *I* increases from 2.8 to 8.4 µA from bottom to top. **d.** Corresponding maximum values of $|\partial V/\partial \Phi|$ versus the biasing current *I* of sample **D4**.

Figure 4 | **a.** Voltage modulation of sample **D2** as a function of magnetic field at $T = 2.4$ K. **b.** Magnetic field dependence of $I_c$ of sample **D2** at $T = 2.4$ K. Red solid line represents fitting by (3).

Figure 1

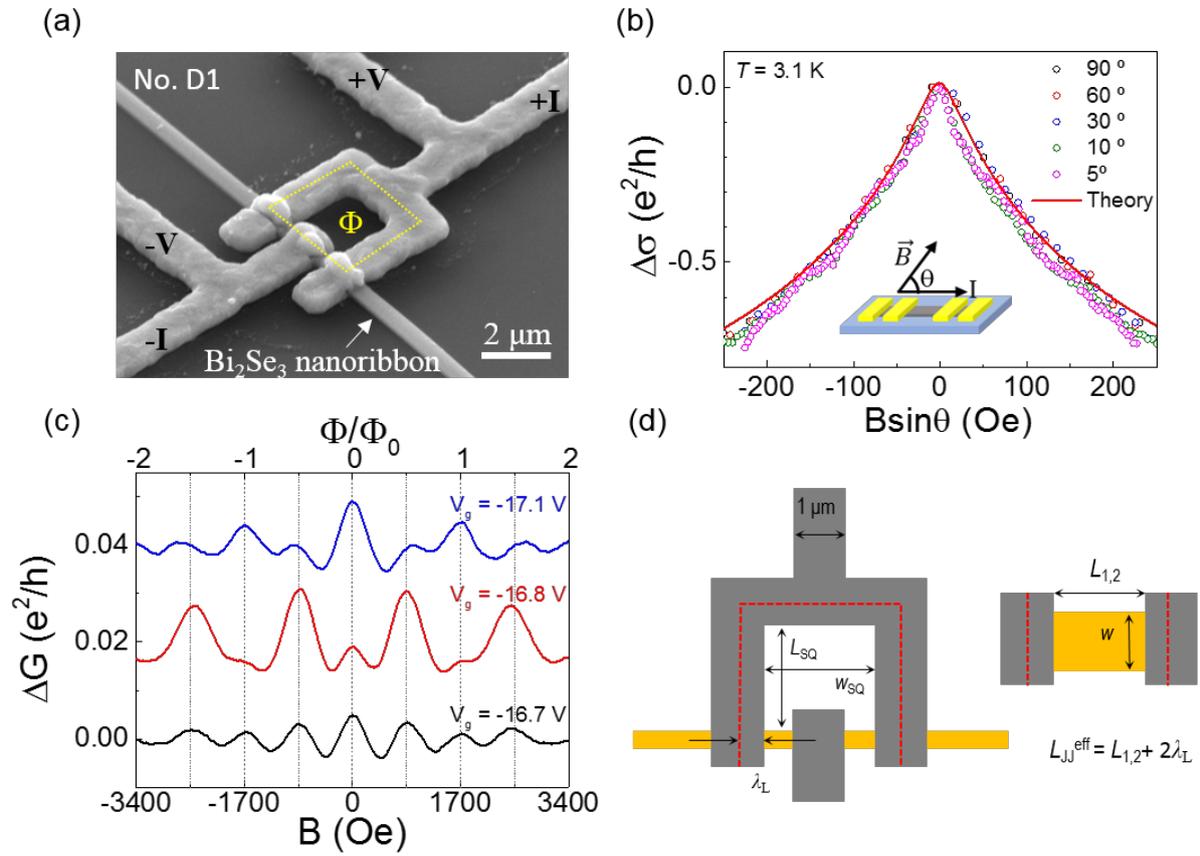

Figure 2

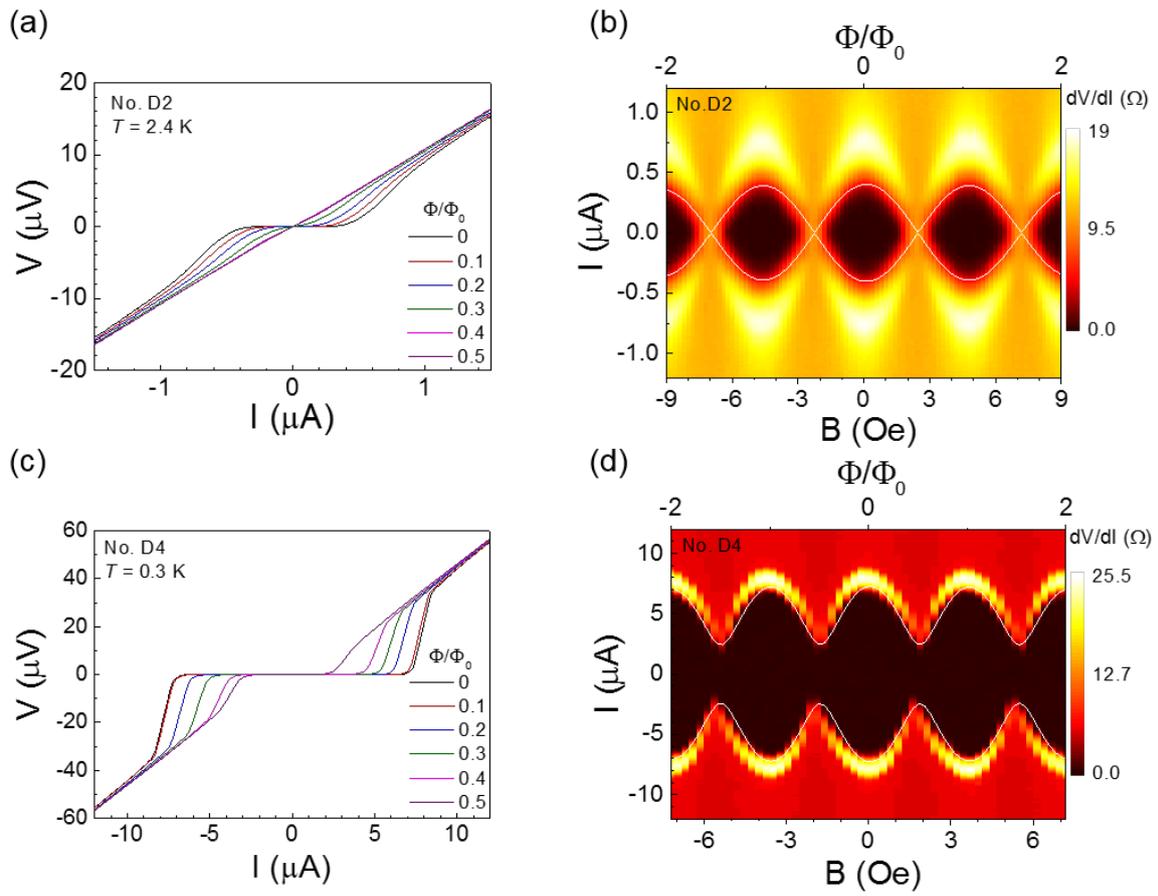

Figure 3

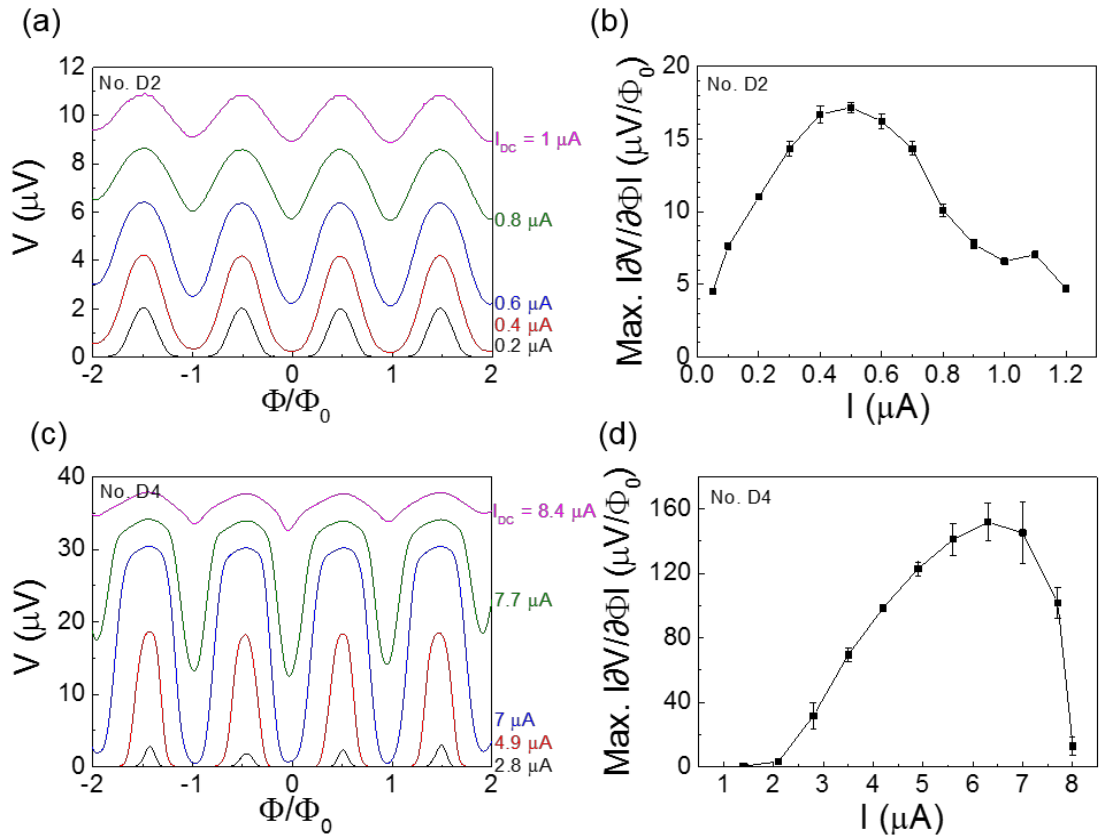

Figure 4

(a) 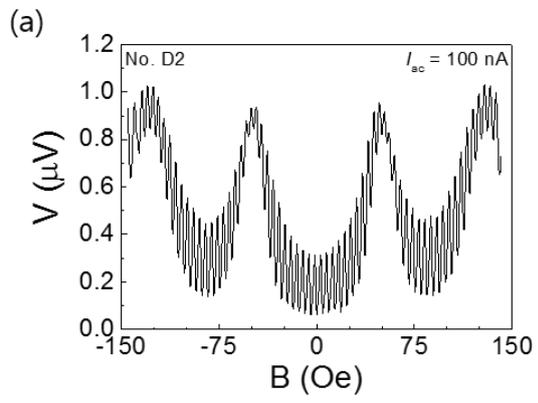 (b) 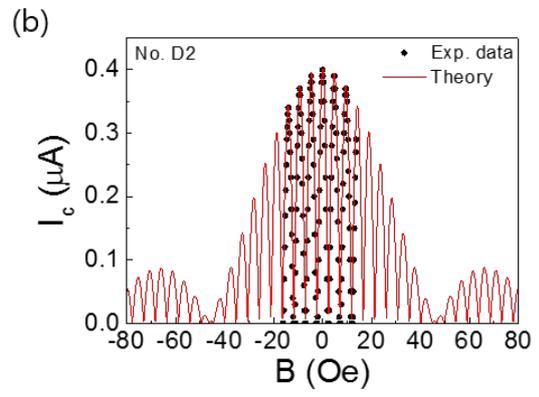